\documentclass[prl,aps,floatfix,showpacs,tightenlines,twocolumn,superscriptaddress,amsmath,amssymb]{revtex4}

\usepackage{graphicx}
\usepackage{dcolumn}
\usepackage{bm}
\usepackage{epsfig}

\begin{document}

\title{Building Gaussian Cluster States by Linear Optics}

\author{Peter van Loock}
\email{vanloock@nii.ac.jp} \affiliation{National Institute of
Informatics, 2-1-2 Hitotsubashi, Chiyoda-ku, Tokyo 101-8430,
Japan}
\author{Christian Weedbrook}
\affiliation{Department of Physics, The University of Queensland,
Brisbane, Queensland 4072, Australia}
\author{Mile Gu}
\affiliation{Department of Physics, The University of Queensland,
Brisbane, Queensland 4072, Australia}

\begin{abstract}
The linear optical creation of Gaussian cluster states, a
potential resource for universal quantum computation, is
investigated. We show that for any Gaussian cluster state, the
canonical generation scheme in terms of QND-type interactions, can
be entirely replaced by off-line squeezers and beam splitters.
Moreover, we find that, in terms of squeezing resources, the
canonical states are rather wasteful and we propose a systematic
way to create cheaper states. As an application, we consider
Gaussian cluster computation in multiple-rail encoding. This
encoding may reduce errors due to finite squeezing, even when the
extra rails are achieved through off-line squeezing and linear
optics.
\end{abstract}


\maketitle

{\it Introduction.---}The cluster-state model for quantum
computation \cite{Raussendorf} is a conceptually interesting
alternative to the more conventional circuit model
\cite{NielsenChuang}. Once a suitable multi-party entangled
cluster state has been prepared, universal quantum gates can be
effected through the cluster via only single-party projective
measurements and feedforward. Though originally based upon qubits,
the cluster-state model can be also applied to other
discrete-variable systems (qudits)
as well as to continuous quantum variables \cite{clusterPRL}.

Linear optics represents one of the most practical approaches to
the realization of quantum information protocols, both for
discrete-variable (DV) \cite{Kok2005} and continuous-variable (CV)
implementations \cite {cvRMP2005}. In the DV case, efficient
entangling gates cannot be achieved with single photons and linear
optics. Nonetheless, probabilistic gates can be applied off-line
to an entangled multi-photon state that eventually serves as a
resource for the on-line computation \cite{KLM01}. A similar
approach uses DV photonic cluster states, leading to a significant
reduction in the resource consumption
\cite{Nielsen2004,Browne2005}. However, the generation of the
optical cluster states remains highly probabilistic in this case.

Although up to six-qubit single-photon cluster states have been
created via postselection using nonlinear and linear optics
\cite{Walther2005,Lu2006}, a possible deterministic, unconditional
realization of optical cluster states would be based on continuous
variables. Here, the resources are squeezed states of light and
the Gaussian cluster states may be created via quadratic quantum
nondemolition (QND) interactions \cite{Zhang06}. These
interactions, however, cannot be realized through beam splitters
alone. Additional ``on-line'' squeezers are needed for every
single link of the cluster state, again rendering the mechanism
for cluster generation rather inefficient with current technology.
Moreover, the squeezing of the resource states will always be
finite, inevitably resulting in errors in the cluster computation.
In this paper, we will address both issues: the avoidance of
on-line squeezing in cluster-state generation and the reduction of
finite-squeezing induced errors in cluster-state computation.


We will show that for any Gaussian cluster state, the canonical
generation scheme in terms of QND-type interactions, can be
entirely replaced by off-line squeezers and beam splitters.
Moreover, we propose a systematic way on how to build alternative
cluster-type states from potentially cheaper squeezing resources
than needed for the canonical states. In any of these
linear-optics schemes, the resource states require correspondingly
more squeezing to compensate for the lack of extra squeezing in
the beam-splitter network that replaces the QND coupling of the
cluster nodes. Nonetheless, the main features of the canonical
QND-made clusters can be preserved. As an example, we consider
Gaussian cluster computation in multiple-rail encoding. This
encoding may reduce errors caused by finite squeezing. We will see
that multiple-rail encoding with linear-optics-made clusters,
though requiring supposedly more off-line squeezing for the extra
links of the larger clusters, can still lead to the same error
reduction as for the canonical cluster states.

We define cluster-type states as those multi-mode Gaussian states
for which certain quadrature correlations become perfect in the
limit of infinite squeezing \cite{Zhang06},
\begin{equation}\label{graphcorr}
\left(\hat p_a - \sum_{b\in N_a} \hat x_b\right)\rightarrow
0\,,\quad a\in G\,.
\end{equation}
Perfect correlations uniquely define the corresponding graph state
(for the DV case, see \cite{Hein}).
Here we use the dimensionless ``position'' and ``momentum''
operators, $\hat x$ and $\hat p$, corresponding to the quadratures
of an optical mode with annihilation operator $\hat a = \hat x + i
\hat p$. The modes $a \in G$ correspond to the vertices of the
graph, while the modes $b\in N_a$ are the nearest neighbors of
mode $a$.

{\it Canonical cluster states via linear optics.---}The canonical
way to build CV cluster states would be to send a number of
single-mode squeezed states through a corresponding network of QND
gates \cite{Zhang06,clusterPRL}. Each individual QND gate could be
realized via two beam splitters and a pair of on-line squeezers
\cite{Braunstein05}. However, by including the initial single-mode
squeezers into the QND network, the resulting total circuit
corresponds to a big quadratic Hamiltonian applied to a number of
vacuum modes. This total transformation can be decomposed into a
linear-optics circuit followed by single-mode squeezers and a
second linear-optics circuit \cite{Braunstein05}, where the first
linear-optics circuit has no effect on the vacuum modes.
Eventually one has just one linear circuit applied to a number of
single-mode squeezed states; in principle, this works for any
cluster (or graph) state. Let us explain this in a little more
detail.

The canonical generation of CV cluster states \cite{Zhang06} from
momentum-squeezed vacuum modes via QND-type interactions can be
described by
$\hat x_a' = \hat x_a$ and
$\hat p_a' = \hat p_a + \sum_{b\in N_a} \hat x_b$,
where
$\hat x_a = e^{+r} \hat x_a^{(0)}$ and
$\hat p_a = e^{-r} \hat p_a^{(0)}$,
$\forall\, a \in G$, with vacuum modes labeled by superscript
$(0)$. Note that, according to Eq.~(\ref{graphcorr}), the
canonical cluster states become perfectly correlated for
$r\to\infty$. We have the following linear Bogoliubov (LUBO)
transformation,
\begin{eqnarray}
\hat a_a' = \hat x_a' + i \hat p_a'
\stackrel{!}{=} \sum_{l\in G} A_{al}\, \hat a_l^{(0)} + B_{al}\,
\hat a_l^{(0)\dagger}\,.
\end{eqnarray}
Using $\hat a_l^{(0)}= \hat x_l^{(0)} + i \hat p_l^{(0)}$, we can
extract the LUBO matrix elements for canonical cluster generation:
$A_{aa}=\cosh r$,
$B_{aa}=\sinh r$,
$A_{ab}=B_{ab}= (i/2) e^{+r}$
($\forall\,b\in N_a$), and
$A_{al}=B_{al}=0$
[$\forall\,l\not\in N_a\, (l\neq a)$].
Now the input modes of the LUBO transformation are vacuum modes
instead of the squeezed input modes of the QND network. The next
step is to decompose this LUBO transformation into a linear-optics
circuit $V^\dagger$, a set of single-mode squeezers, and another
linear-optics circuit $U$ \cite{Braunstein05}.
Using the singular value decomposition for $A$ and $B$, we have to
satisfy
$V=A^\dagger U A_D^{-1}
=B^T U^* B_D^{-1}$ and
$U=A V A_D^{-1}
=B V^* B_D^{-1}$
for the unitary matrices $U$ and $V$. Here $A_D^2$ is the
diagonalized version of $A A^\dagger$ and $A^\dagger A$;
similarly, we use $B_D^2$ for $B B^\dagger$ and $(B^\dagger B)^T$.
The singular-value matrix equations lead to the conditions for
$U$,
\begin{equation}\label{simplecriteria2}
{\rm Im}U_{al} - C_l(r)\sum_{b\in N_a} {\rm Re}U_{bl} = 0\,, \quad
\forall\, a,l \in G\,,
\end{equation}
with
\begin{equation}\label{constraints}
{\rm Re}U_{al}\left[ D_l(r) - M_{N_a}\right] - \sum_{b\in
N_a}\sum_{k\in N_b,k\neq a} {\rm Re}U_{kl} = 0\,,
\end{equation}
where
\begin{equation} C_l(r)\equiv \frac{\frac{e^{+r}}{2}
\left(\sqrt{\lambda_l^A} +
\sqrt{\lambda_l^B}\right)}{\sqrt{\lambda_l^A}\sinh r +
\sqrt{\lambda_l^B}\cosh r}\,,
\end{equation}
$D_l(r)\equiv (\lambda_l^B\cosh^2 r - \lambda_l^A\sinh^2
r)/[e^{+2r}(\lambda_l^A - \lambda_l^B)/4]$, and ${\rm
Re}U_{al}\equiv {\rm Re}(U_{al})$, etc.; $M_{N_a}$ represents the
number of nearest neighbors of mode $a$. The double sum in
Eq.~(\ref{constraints}) sums over all second neighbors of mode $a$
(including multiple counting of identical neighbors of the nearest
neighbors of mode $a$). The expressions $\sqrt{\lambda_l^A}$ are
the singular values of $A$ and similarly for $B$.
Equation~(\ref{simplecriteria2}) can be incorporated into the
column vectors of $U$,
\begin{eqnarray}\label{vectors2}
\vec u_l \equiv \left( \begin{array}{c} \alpha_{1l} + i C_l(r)
\sum_{b\in N_1}\alpha_{bl} \\
\alpha_{2l} + i C_l(r)\sum_{b\in N_2}\alpha_{bl} \\
\cdots\\
\alpha_{Nl} + i C_l(r)\sum_{b\in N_N}\alpha_{bl}
\end{array} \right),
\end{eqnarray}
using ${\rm Re}U_{kl}\equiv \alpha_{kl}$, with the constraints of
Eq.~(\ref{constraints}), and $\sum_l (\vec u_l)_k (\vec
u_l)^*_{k'} = \delta_{k k'}$ for unitarity. These conditions
automatically satisfy $A=U A_D V^\dagger$ and $B=U B_D V^T$, where
the diagonal matrices $A_D$ and $B_D$ contain the corresponding
singular values. Thus, we effectively constructed a linear-optics
circuit $U$ that {\it exactly} outputs the canonical cluster
states when applied to the {\it off-line} squeezed input modes
with squeezed quadratures $\hat x_l|\hat p_l =
\left(\sqrt{\lambda_l^A} \pm \sqrt{\lambda_l^B}\right)\,\hat
x_l^{(0)}|\hat p_l^{(0)} \equiv e^{\pm R_l}\,\hat x_l^{(0)}|\hat
p_l^{(0)}$.
The vacuum modes, $\hat a_l^{(0)}= \hat x_l^{(0)} + i \hat
p_l^{(0)}$, are ``redefined'' vacuum modes after the first
linear-optics circuit $V^\dagger$ that has no effect.

\begin{figure}[t]
\centering
\includegraphics[width=8cm]{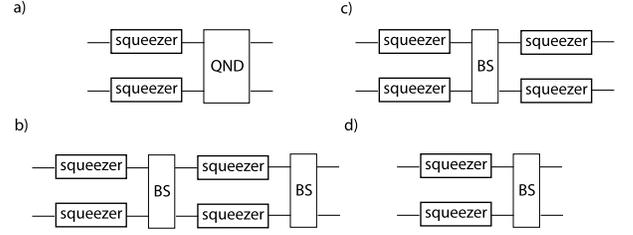}
\caption{Different generation schemes for a canonical two-mode
cluster state. The input modes are always vacuum modes; a)
canonical generation via QND interaction; b) QND gate replaced by
a beam splitter (BS), two on-line squeezers, and another beam
splitter \cite{Braunstein05}; c) generation of a standard two-mode
squeezed state (via two squeezers and a beam splitter) plus two
local squeezers \cite{Duan00}; d) simplest generation scheme with
the QND gate replaced by a beam splitter.} \label{fig1}
\end{figure}

Let us consider the example of the canonical two-mode cluster
state. It corresponds to two momentum-squeezed modes (squeezed by
$r$) coupled through a quadratic QND gate $e^{2i \hat x
\otimes\hat x}$, see Fig.\ref{fig1}a. In this case, we have
$\lambda_l^A\equiv \cosh^2 r + e^{+2r}/4$, $\lambda_l^B\equiv
\sinh^2 r + e^{+2r}/4$, $C_1(r)=C_2(r)\equiv C^{-1}(r)$, and
$D_l(r)\equiv 1$. Thus, in the equivalent linear-optics scheme
(Fig.\ref{fig1}d), two equally squeezed modes
are combined at a beam
splitter described by $U$ with column vectors as in
Eq.~(\ref{vectors2}), $l=1,2$, $N=2$; the constraints in
Eq.~(\ref{constraints}) are always satisfied.

A possible solution for $U$, choosing $\alpha_{12}=\alpha_{21}=0$,
is
\begin{eqnarray}\label{standardtwomodes}
U = \frac{1}{\sqrt{1+C^2(r)}}
\left( \begin{array}{cc} C(r) & i \\
i & C(r)
\end{array} \right)\,.
\end{eqnarray}
Each input mode is momentum-squeezed with $R_l=\ln
\left(\sqrt{\lambda_l^A}+\sqrt{\lambda_l^B}\right)\equiv R$. For
$r=0$, we obtain $\lambda_l^A\equiv 5/4$ and $\lambda_l^B\equiv
1/4$, and thus $R=\ln[(1+\sqrt{5})/2]$. This is the residual
squeezing coming from the QND gate, which is now additionally
applied off-line before the beam splitter. Note that the QND gate
here is not simply replaced by a beam splitter followed by two
single-mode squeezers and another beam splitter
\cite{Braunstein05}, all together applied to two initial squeezed
states (Fig.\ref{fig1}b). Instead, the circuit is further
simplified, and only {\it one} pair of squeezers followed by {\it
one} beam splitter operation act upon initial vacuum modes.
However, in order to produce exactly the canonical state, this
circuit is rather wasteful in terms of squeezing resources. The
most economical two-mode entangled state is a standard two-mode
squeezed state (TMSS) built from two single-mode squeezed states
with a 50/50 beam splitter.
Like any pure Gaussian two-mode state \cite{Duan00}, also the
canonical two-mode cluster state can be obtained from a TMSS via
local Gaussian transformations, including local squeezers
(Fig.\ref{fig1}c).

Comparing resources, for example, an excess noise of one vacuum
unit in the quadrature correlations $\hat p_1 - \hat x_2$ and
$\hat p_2 - \hat x_1$ can be achieved with a TMSS built from two 3
dB squeezed states (see below). The canonical two-mode cluster
state with these correlations ($r=0$), would require two 4.18 dB
squeezed states combined at the asymmetric beam splitter in
Eq.~(\ref{standardtwomodes}).
In general, the canonical $N$-mode cluster states are always
biased in $x$ and $p$ \cite{Bowen03}, $\langle\hat x_a^2\rangle =
e^{+2r} \neq \langle\hat p_a^2\rangle = e^{-2r} + M_{N_a}
e^{+2r}$, unnecessarily requiring extra {\it local squeezing} to
achieve a certain degree of correlations and entanglement; they
cannot be obtained from $N$-mode pure Gaussian states in standard
form \cite{Adesso06a}
without the use of local squeezers.
In the following, we shall consider a whole family of states
producible via linear optics and exhibiting cluster-type
correlations. This family will include states cheaper than the
canonical cluster states.

{\it Cluster-type states via linear optics.---}In order to create
states with correlations as in Eq.~(\ref{graphcorr}) via linear
optics, we consider $p$-squeezed input modes,
$\hat a_l = e^{+R_l} \hat x_l^{(0)} + i e^{-R_l} \hat p_l^{(0)}$,
and a general linear-optics transformation,
$\hat a_k'= \sum_{l} U_{kl}\, \hat a_l$,
with a unitary matrix $U$. Using $\hat a_k' = \hat x_k' +i \hat
p_k'$, one obtains the output quadrature operators, and
in order to satisfy the correlations in Eq.~(\ref{graphcorr}), we
assume the large noise terms (those proportional to $e^{+R_l}$)
cancel.
This is possible if and only if
\begin{equation}\label{simplecriteria}
{\rm Im}U_{al} - \sum_{b\in N_a} {\rm Re}U_{bl} = 0\,, \quad
\forall\, a,l \in G\,.
\end{equation}
After inserting these conditions into the excess noise terms
(those that vanish for infinite squeezing $R_l\to\infty$), we find
that every quadrature correlation of Eq.~(\ref{graphcorr}) labeled
by $a$ has an excess noise variance
\begin{equation}\label{excessnoise}
\sum_l \left[{\rm Re}U_{al}\left( 1+M_{N_a}\right) + \sum_{b\in
N_a}\sum_{k\in N_b,k\neq a} {\rm Re}U_{kl}\right]^2 e^{-2R_l}\,,
\end{equation}
times one unit of vacuum quadrature noise ($1/4$ in our scales)
which we omit in the following. These excess noises will lead to
the errors in the cluster computation \cite{clusterPRL}. Let us
consider two-mode cluster-type states.

Look at the simple circuit described by
\begin{eqnarray}\label{2modecircuit}
U =
\left( \begin{array}{cc} 1 & 0 \\
0 & i
\end{array} \right)\!\!\!
\left( \begin{array}{cc} \frac{1}{\sqrt{2}} & \frac{1}{\sqrt{2}} \\
\frac{1}{\sqrt{2}} & -\frac{1}{\sqrt{2}}
\end{array} \right)\!\!\!
\left( \begin{array}{cc} 1 & 0 \\
0 & i
\end{array} \right) =
\left( \begin{array}{cc} \frac{1}{\sqrt{2}} & \frac{i}{\sqrt{2}} \\
\frac{i}{\sqrt{2}} & \frac{1}{\sqrt{2}}
\end{array} \right)\!\!,
\end{eqnarray}
applied to two momentum-squeezed modes. The circuit is a 50/50
beam splitter with Fourier transforms of mode 2 before and after
the beam splitter. It clearly satisfies the conditions in
Eq.~(\ref{simplecriteria}).
The resulting state is the standard TMSS up to a local Fourier
transform of mode 2. No extra local squeezers are needed to obtain
this cluster-type state from standard two-mode entanglement.
According to Eq.~(\ref{excessnoise}), the excess noises in the
cluster-type correlations are $2e^{-2R_1}$ in $\hat p_1 - \hat
x_2$ and $2e^{-2R_2}$ in $\hat p_2 - \hat x_1$, corresponding to
the optimal entanglement for a given input squeezing. Note that
the solutions in Eq.~(\ref{2modecircuit}) and
Eq.~(\ref{standardtwomodes}) coincide for $r\to\infty$ when
$C(r)\to 1$.





\begin{figure}[b]
\centering
\includegraphics[width=8cm]{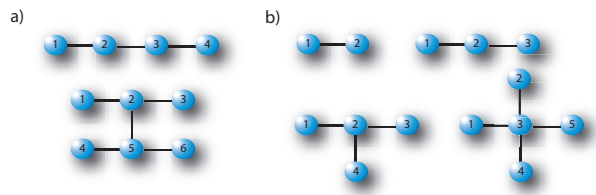}
\caption{Examples of cluster-type states producible via linear
optics; each optical mode/cluster node $a$ is represented by a
vector $\vec{\alpha}_a$ (see text): a) linear 4-mode cluster,
nonlinear 6-mode cluster; b) ``EPR/GHZ-type'' \cite{PvLPRL00}
clusters: linear 2 and 3-mode clusters, T-shape and cross-shape
clusters.} \label{fig2}
\end{figure}

Incorporating Eq.~(\ref{simplecriteria}) into the column vectors
of $U$ leads to
\begin{eqnarray}\label{vectors}
\vec u_l \equiv
\left( \begin{array}{c} \alpha_{1l} + i \sum_{b\in N_1}\alpha_{bl} \\
\alpha_{2l} + i \sum_{b\in N_2}\alpha_{bl} \\
\cdots\\
\alpha_{Nl} + i \sum_{b\in N_N}\alpha_{bl}
\end{array} \right),
\end{eqnarray}
with $\sum_l (\vec u_l)_k (\vec u_l)^*_{k'} = \delta_{k k'}$ for
unitarity and again ${\rm Re}U_{kl}\equiv \alpha_{kl}$. Using this
formalism, for any given graph state, one obtains a set of
geometrical conditions for $N$ real vectors $\vec \alpha_k^T
\equiv (\alpha_{k1},\alpha_{k2},...,\alpha_{kN})$.
For instance, a
linear 4-mode cluster state has
$\vec \alpha_1 \vec \alpha_3 = \vec \alpha_2 \vec \alpha_4 =
-1/5$,
$\vec \alpha_1 \vec \alpha_2 = \vec \alpha_1 \vec \alpha_4 = \vec
\alpha_2 \vec \alpha_3 = \vec \alpha_3 \vec \alpha_4 = 0$,
$||\vec \alpha_1 || = ||\vec \alpha_4 || = 3/5$, and
$||\vec \alpha_2 || = ||\vec \alpha_3 || = 2/5$,
achievable, for example, with
$\vec \alpha_1^T = (1/\sqrt{2},1/\sqrt{10},0,0)$,
$\vec \alpha_2^T = (0,0,-2/\sqrt{10},0)$,
$\vec \alpha_3^T = (0,-2/\sqrt{10},0,0)$, and
$\vec \alpha_4^T = (0,0,1/\sqrt{10},-1/\sqrt{2})$.
This, together with Eq.~(\ref{vectors}), where we set $N=4$,
defines a possible solution $U$. In general, any 4-mode
transformation $U$ can be decomposed into a network of
$4(4-1)/2=6$ beam splitters \cite{Reck}. However, for the linear
4-mode cluster,
there is a minimal decomposition of $U$ into a network of only
three beam splitters $U = F_4 S_{12} F_1^\dagger
B_{34}^{+}(1/\sqrt{2}) B_{12}^{+}(-1/\sqrt{2})
B_{23}^{-}(1/\sqrt{5}) F_3^\dagger F_4^\dagger$. Here, $F_k$
(Fourier transform of mode $k$) is the 4-mode identity matrix
$I_4$ except $(F_k)_{kk}=i$, $B_{kl}^{\pm}(t)$ (beam splitter
transformation of modes $k$ and $l$) equals $I_4$ except
$(B_{kl}^{\pm})_{kk}=t$, $(B_{kl}^{\pm})_{kl}=\sqrt{1-t^2}$,
$(B_{kl}^{\pm})_{lk}=\pm \sqrt{1-t^2}$, and
$(B_{kl}^{\pm})_{ll}=\mp t$, and the matrix $S_{12}$ swaps modes 1
and 2.

\begin{figure}[t]
\centering
\includegraphics[width=8cm]{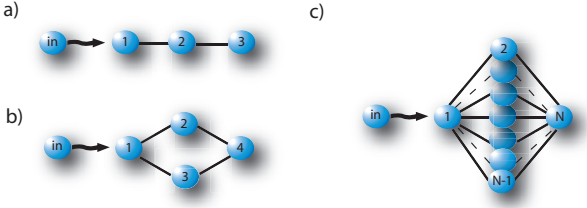}
\caption{Error propagation when an input state is attached to and
teleported through a) a linear 3-mode cluster; b) a 4-mode diamond
cluster; c) an $N$-mode multiple-rail cluster; all clusters are
generated via off-line squeezing and linear optics.} \label{fig3}
\end{figure}

Similarly, one can find linear-optics solutions for other
cluster-type states (Fig.\ref{fig2}).
The example of a nonlinear 6-mode cluster in Fig.\ref{fig2}a (a
potential resource for {\it two-mode} evolutions) leads to two
orthogonal subspaces $\{\vec \alpha_k, \vec \alpha_l,\vec
\alpha_m\}$ with $\{k,l,m \}=\{1,3,5 \}$ and $\{k,l,m \}=\{6,4,2
\}$, where $\vec \alpha_k \vec \alpha_l = -3/10$, $\vec \alpha_k
\vec \alpha_m = \vec \alpha_l \vec \alpha_m = -1/10$, and $||\vec
\alpha_k || = ||\vec \alpha_l || = 7/10$, $||\vec \alpha_m || =
3/10$. A possible solution is $(1/\sqrt{5},1/\sqrt{2},0,0,0,0)$,
$(1/\sqrt{5},-1/\sqrt{2},0,0,0,0)$, $(-1/\sqrt{20},0,1/2,0,0,0)$
for vectors $\{1,3,5 \}$, respectively, and also
$(0,0,0,-1/\sqrt{20},0,1/2)$, $(0,0,0,1/\sqrt{5},1/\sqrt{2},0)$,
$(0,0,0,1/\sqrt{5},-1/\sqrt{2},0)$ for $\{2,4,6 \}$. In general,
using the geometrical conditions for $N$ $N$-dimensional vectors,
solutions can be recursively constructed for any graph state:
start with $\vec \alpha_1^T = (\alpha_{11},0,0,\hdots)$ determined
by the norm $||\vec \alpha_1 ||$, then $\vec \alpha_2^T =
(\alpha_{21},\alpha_{22},0,\hdots)$ given by the overlap $\vec
\alpha_1 \vec \alpha_2$ and the norm $||\vec \alpha_2 ||$, $\vec
\alpha_3^T = (\alpha_{31},\alpha_{32},\alpha_{33},\hdots)$ via
$\vec \alpha_1 \vec \alpha_3$, $\vec \alpha_2 \vec \alpha_3$, and
$||\vec \alpha_3 ||$, etc.



{\it Redundant encoding for error filtration.---}As an
application, we consider multiple-rail encoding in a Gaussian
cluster computation protocol \cite{clusterPRL}. Look at the
diamond cluster state in Fig.\ref{fig3}b. In this case, the vector
conditions become $\vec \alpha_1 \vec \alpha_2 = \vec \alpha_1
\vec \alpha_3 = \vec \alpha_2 \vec \alpha_4 = \vec \alpha_3 \vec
\alpha_4 = 0$, $\vec \alpha_2 \vec \alpha_3 = \vec \alpha_1 \vec
\alpha_4 = - 2/5$, and $||\vec \alpha_k || \equiv 3/5$. This can
be satisfied, for example, via
$\vec \alpha_1^T = (1/\sqrt{3},-2/\sqrt{15},0,0)$,
$\vec \alpha_2^T = (0,0,1/\sqrt{3},-2/\sqrt{15})$,
$\vec \alpha_3^T = (0,0,0,\sqrt{3/5})$, and
$\vec \alpha_4^T = (0,\sqrt{3/5},0,0)$.
A simple cluster protocol would be to teleport an input mode onto
mode 1 of the diamond state and from there to mode 4
(Fig.\ref{fig3}b). For simplicity, we assume that the input mode
is attached to mode 1 through the QND gate $e^{2i \hat x_{\rm in}
\otimes\hat x_1}$ such that $\hat p_1\to \hat p_1 + \hat x_{\rm
in}$ and $\hat p_{\rm in}\to \hat p_{\rm in} + \hat x_1$; $\hat
x_1$ and $\hat x_{\rm in}$ remain unchanged. Now $p$-measurements
of the input mode and modes 1 through 3 (with results $s_{\rm
in}$, $s_{1\cdots 3}$) would transform mode 4 into the input state
up to some known corrections and excess noise coming from the
imperfect diamond state. After the correction operation, $\hat
F^\dagger \hat X(-s_{\rm in}) \hat F^\dagger \hat X(-s_1) \hat
F^\dagger \hat X(-(s_2+s_3)/2)$, with an inverse Fourier transform
operator $\hat F^\dagger$ acting as $\hat x\to \hat p$ and $\hat
p\to -\hat x$, and $x$-displacements $\hat X(s)$ such that $\hat x
\to \hat x + s$, mode 4 is described by $\hat x_{\rm out} = \hat
x_{\rm in} + (\hat p_1 - \hat x_2 - \hat x_3) - (\hat p_4 - \hat
x_2 - \hat x_3)$ and $\hat p_{\rm out} = \hat p_{\rm in} - (\hat
p_2 - \hat x_1 - \hat x_4)/2 - (\hat p_3 - \hat x_1 - \hat
x_4)/2$. Here, $\hat p_{\rm in}$ and $\hat p_1$ correspond to the
momentum operators of the input mode and mode 1 before the QND
coupling.

We see that the imperfect teleportation fidelities depend on the
correlations of the diamond cluster, Eq.~(\ref{graphcorr}) with
Fig.\ref{fig3}b. In the corresponding excess noise variances,
Eq.~(\ref{excessnoise}), let us consider only the effect of finite
$R_3$ and $R_4$, assuming $R_1\gg 1$ and $R_2\gg 1$. This leads to
negligible excess noises, $a=1$ and $a=4$, in $\hat x_{\rm out}$,
because the vectors $\vec \alpha_1$ and $\vec \alpha_4$ live in
the two-dimensional subspace $l=1,2$. In $\hat p_{\rm out}$,
according to Eq.~(\ref{excessnoise}) with $a=2$ and $a=3$, we find
a total excess noise of $(13\, e^{-2R_3} + 5\, e^{-2R_4})/12= 3\,
e^{-2R}/2$, assuming $R_3=R_4=R$, which is {\it half} the excess
noise compared to {\it any} linear 3-mode cluster protocol
(Fig.\ref{fig3}a) with two modes highly squeezed and one finitely
squeezed by $R$; the generality of this result can be easily
proven using Eq.~(\ref{excessnoise}) and Eq.~(\ref{vectors}) for
the 3-mode protocol, with $\hat p_{\rm out} = \hat p_{\rm in} -
(\hat p_2 - \hat x_1 - \hat x_3)$ and $\hat x_{\rm out} = \hat
x_{\rm in} + (\hat p_1 - \hat x_2) - (\hat p_3 - \hat x_2) \approx
\hat x_{\rm in}$. More generally, we obtain a $p$ excess noise of
$3\, e^{-2R}/m$ for multiple-rail encoding (Fig.\ref{fig3}c),
where $m$ is the number of intermediate nodes (number of rails)
between mode 1 and the output mode \cite{footnote1}. As a result,
the excess noise in cluster computation may be reduced by
teleporting the input through multiple paths. Remarkably, the
linear-optics-made clusters achieve the same error reduction as
obtainable for QND-made clusters with QND gates freely available.

{\it Conclusion.---}We showed that any Gaussian cluster state can
be built via off-line squeezing and linear optics without QND
couplings. Simple vector conditions lead to potentially cheaper
cluster-type states. In multiple-rail-encoded cluster computation,
the same error reduction as for the QND-made clusters can be
achieved. Our results pave the way for experimental realizations
of small-scale cluster computation with continuous variables.

PvL acknowledges the MIC in Japan, CW and MG thank the Australian
Research Council for support.


\begin{thebibliography}{99}
\bibitem{Raussendorf} R.\ Raussendorf and H.~J.\ Briegel,
Phys.\ Rev.\ Lett.\ {\bf 86}, 5188 (2001).
\bibitem{NielsenChuang} M.\ A.\ Nielsen and I.\ L.\ Chuang,
{\it Quantum Computation and Quantum Information} (Cambridge
University Press, 2000).
\bibitem{clusterPRL} N.\ C.\ Menicucci {\it et al.},
Phys.\ Rev.\ Lett.\ {\bf 97}, 110501 (2006).
\bibitem{Kok2005} P.\ Kok {\it et al.},
eprint: quant-ph/0512071 (2006).
\bibitem{cvRMP2005} S.~L.\ Braunstein and P.\ van Loock,
Rev.\ Mod.\ Phys.\ {\bf 77}, 513 (2005).
\bibitem{KLM01} E.\ Knill, R.\ Laflamme, and G.\ J.\ Milburn,
Nature {\bf 409}, 46 (2001).
\bibitem{Nielsen2004} M.\ A.\ Nielsen,
Phys.\ Rev.\ Lett.\ {\bf 93}, 040503 (2004).
\bibitem{Browne2005} D.\ E.\ Browne and T.\ Rudolph,
Phys.\ Rev.\ Lett.\ {\bf 95}, 010501 (2005).
\bibitem{Walther2005} P.\ Walther {\it et al.},
Nature {\bf 434}, 169 (2005).
\bibitem{Lu2006} C.~Y. Lu {\it et al.},
eprint: quant-ph/0609130 (2006).
\bibitem{Zhang06} J.\ Zhang and S.~L.\ Braunstein,
Phys.\ Rev.\ A {\bf 73}, 032318 (2006).
\bibitem{Hein} M.\ Hein, J.\ Eisert, and H.\ J.\ Briegel,
Phys.\ Rev.\ A {\bf 69}, 062311 (2004).
\bibitem{Braunstein05} S.~L.\ Braunstein,
Phys.\ Rev.\ A {\bf 71}, 055801 (2005).
\bibitem{Duan00} L.-M.\ Duan {\it et al.},
Phys.\ Rev.\ Lett.\ {\bf 84}, 2722 (2000).
\bibitem{Bowen03} W.\ P.\ Bowen, P.\ K.\ Lam, and T.\ C.\ Ralph,
J.\ Mod.\ Opt.\ {\bf 50}, 801 (2003).
\bibitem{Adesso06a} G.\ Adesso,
Phys.\ Rev.\ Lett.\ {\bf 97}, 130502 (2006).
\bibitem{Reck} M.\ Reck {\it et al.},
Phys.\ Rev.\ Lett.\ {\bf 73}, 58 (1994).
\bibitem{PvLPRL00} P.\ van Loock and S.\ L.\ Braunstein,
Phys.\ Rev.\ Lett.\ {\bf 84}, 3482 (2000).
\bibitem{footnote1} for the general case, the vector conditions
are $\vec \alpha_1 \vec \alpha_N = (m - 2 m^2)/(4 m^2 - 1)$, $\vec
\alpha_k \vec \alpha_l = (2 - 4 m)/(4 m^2 - 1)$, ($k,l\neq 1,N$),
$\vec \alpha_1 \vec \alpha_k = \vec \alpha_N \vec \alpha_k = 0$,
$||\vec \alpha_1 || = ||\vec \alpha_N || = (2 m^2 + m - 1)/(4 m^2
- 1)$, and $||\vec \alpha_k || = (4 m^2 - 4 m + 1)/(4 m^2 - 1)$.
\end{thebibliography}
\end{document}